\documentclass[showkeys,superscriptaddress,10pt,balancelastpage,twocolumn]{revtex4}%
\usepackage[verbose,colorlinks,hyperindex,breaklinks=true,pdfusetitle,citecolor=blue,urlcolor=cyan]{hyperref}
\usepackage{amsfonts}
\usepackage{amsmath}
\usepackage{amssymb}
\usepackage{color}
\usepackage{graphicx}

\begin{document}
\title{Exponential localization of moving-end mirror in an optomechanical system}
\author{Kashif Ammar Yasir\footnote{\textcolor{red} {kashif\_ammar@yahoo.com}}}
\affiliation{Department of Electronics, Quaid-i-Azam University, 45320, Islamabad, Pakistan.}
\author{Muhammad Ayub}
\affiliation{Department of Electronics, Quaid-i-Azam University, 45320, Islamabad, Pakistan.}
\affiliation{LINAC Project, PINSTECH, Nilore, 45650, Islamabad, Pakistan.}
\author{Farhan Saif}
\affiliation{Department of Electronics, Quaid-i-Azam University, 45320, Islamabad, Pakistan.}

\begin{abstract}
We discuss the dynamics of moving end mirror of an optomechanical system that consists of a Fabry-Perot cavity 
loaded with dilute condensate and driven by a single mode optical field. It is shown that quantum mechanical phenomenon of 
dynamical localization occurs both in position and momentum space for moving end mirror in the system. The parametric dependencies of 
dynamical localization are discussed. We also provide a set of parameters which makes this phenomenon experimentally feasible.
\end{abstract}

\maketitle

\section{Introduction}
\label{intro}
Cavity-optomechanics deals with the interaction of an optical field in a resonator with confining mirrors~\cite{Kippenberg}. 
Recent experimental advances make it possible to couple cold atoms and Bose-Einstein condensates (BEC), mechanical membrane and 
nano-sphere with the optomechanical systems \cite{Esslinger}. Hence, these hybrid-opto mechanical systems are playground to study 
phenomena related to mirror-field interaction, and atom-field interaction which provide founding principles to develop numerous sensors, and devices in quantum metrology~\cite{saifrtm,saifopt}. 
In opto-mechanics the mechanical effects of light lead to cooling the motion of a movable mirror to its quantum mechanical ground state \cite{CornnellNat2010,TeufelNat2011,CannNat2011}, 
and to observe strong coupling effects~\cite{GroeblacherNat2009,ToefelNat2011a,VerhagenNat2012}. 
In earlier work optomechanical systems were proposed precision measurement in gravitational wave detectors
~\cite{Braginsky} and to measure displacement with large accuracy~\cite{Rugar}, and recently the
development of optomechanical crystals~\cite{EchenfieldNat2012}. Recent theoretical discussions and simulations on bistable behavior of BEC-optomechanical system
\cite{Meystre2010}, high fidelity state transfer \cite{YingPRL2012,SinghPRL2012}, steady-state entanglement of BEC and moving end mirror~\cite{Asjad2011,Asjad2012}, macroscopic tunneling of an optomechanical membrane  
\cite{Buchmann2012} and role reversal between matter-wave and quantized light field are guiding towards new avenues in cavity optomechanics.   
In this submission we report the classical and quantum dynamics in position and momentum space and show that dispersion is suppressed in quantum domain while classical~\cite{Yang,Zhang} 
counterpart increases with time following a power-law. This contrasting behavior is an analog of Anderson localization and termed as dynamical localization or exponential localization~\cite{dynloc}. 
Dynamical localization is studied in periodically driven nonlinear systems \cite{Fishman1982,Blumel1991,Moore1994,Bardroff1995,saifdyn,Schelle2009} and provides quantum mechanical limits on classical diffusion of wave packet. The phenomenon of dynamical localization emerges in the explicitly time dependent system. The single mode laser field excite condensate atoms to momentum side modes which
behave formally like a moving mirror driven by the radiation force of intra-cavity field. The side modes induce phase modulation to the field \cite{Meystre2013}, which provide modulation to the radiation pressure exerted on mirror \cite{nature1988}. In this paper, the model of the system is presented in section~\ref{sec:Model}. 
In section~\ref{sec:Langevin}, we derive the Langevin equation and in section~\ref{sec:NumDL}, 
we explain dynamical localization of the condensates in the system. In section~\ref{sec:ModEffect}, 
we explain dynamical localization as a function of modulation amplitudes. 
\section{The Model}\label{sec:Model}
We consider a Fabry-P\'{e}rot cavity of length $L$ with a moving end mirror driven by a single mode 
optical field of frequency $\omega_{p}$ and BEC with N-two level atoms trapped in an optical lattice potential 
\cite{Nature2008,Science2008}. Moving end mirror has harmonic oscillations with frequency $\omega_{m}$ and 
exhibits Brownian motion in the absence of coupling with radiation pressure.

The Hamiltonian of BEC-optomechanical system is,
\begin{equation}
\hat{H}=\hat{H}_{m}+\hat{H}_{a}+\hat{H}_{T}\label{1},
\end{equation}
where, $\hat{H}_{m}$ describes the intra-cavity field and its coupling to the moving end mirror, $\hat{H_{a}}$ 
accounts 
for the BEC and its coupling with intra-cavity field while, $\hat{H}_{T }$ accounts for noises and 
damping associated with the system. The Hamiltonian $\hat{H}_{m}$ is given as \cite{LawPRA1995},
\begin{equation}
\hat{H}_{m}=\hbar\triangle_{c}\hat{c}^{\dag}\hat{c}+\frac{\hbar\omega_{m}}{2}(\hat{p}^{2}+
\hat{q}^{2})-\xi\hbar\hat{c}^{\dag}\hat{c}\hat{q}-i\hbar\eta(\hat{c}
-\hat{c}^{\dag}),
\end{equation}
where, first term is free energy of the field, $\Delta_{c}=\omega_{c}-\omega_{p}$ is detuning, here, 
$\omega_{c}$ is cavity frequency and $\hat{c}^{\dag}$ ($\hat{c}$) are creation (annihilation) operators 
for intra-cavity field interacting with mirror and condensates and their commutation relation is  
$[\hat{c},\hat{c}^{\dag}]=1$. Second term represents energy of moving end mirror. Here $\hat{q}$ and $\hat{p}$ 
are dimensionless position and momentum operators for moving end mirror, such that, $[\hat{q},\hat{p}]=i$. 
Intra-cavity 
field couples BEC and moving end mirror through radiation pressure. When $n$ photons interacts with the 
surface of mirror 
in cavity round trip time $t=2L/c$, $c$ represents speed of light and $L$ is cavity length,
and transfer $2\hbar k$ momentum to moving end mirror, here $k=\omega_{c}/c$. Therefore, the 
radiation pressure force
$\hat{c}^{\dag}\hat{c}\hbar\omega_{c}/L$ will change the position $\hat q$ of moving end mirror and couples 
moving end mirror
with intra-cavity field. Third term represents this coupling and $\xi=\sqrt{2}(\omega_{c}/L)x_{0}$ is the 
coupling strength where, 
$x_{0}=\sqrt{\hbar/2m\omega_{m}}$, is zero point motion of mechanical mirror of mass $m$. 
Last term gives relation of intra-cavity field and output power $\vert\eta\vert=\sqrt{P\kappa/\hbar\omega_{p}}$, 
where, $P$ is the input laser power and $\kappa$ is cavity decay rate associated with outgoing modes.

Now the Hamiltonian for BEC and intra-cavity field and their coupling is derived by considering quantized 
motion of atoms along 
the cavity 
axis in one dimensional model. We assume that BEC is dilute enough and many body interaction effects are 
ignored~\cite{Meystre2010}. We have

\begin{equation}\label{Ha}
\hat{H}_{a}=\frac{\hbar U_{0}N}{2}\hat{c}^{\dag}\hat{c}+\frac{\hbar\Omega}{2}(\hat{P}^{2}
+\hat{Q}^{2})+\xi_{sm}\hbar\hat{c}^{\dag}\hat{c}\hat{Q}
\end{equation}
here, first term describes energy of field due to the condensate. Second term expresses the energy of the 
condensate in the 
cavity following harmonic motion. Here, $\hat{P}=\frac{i}{\sqrt{2}}(\hat{b}-\hat{b}^{\dag})$ and 
$\hat{Q}=\frac{1}{\sqrt{2}}(\hat{b}+\hat{b}^{\dag})$ are dimensionless momentum and position operators for 
condensate which are related as $[\hat{Q},\hat{P}]=i$ and $\Omega=4\omega_{r}$ where, 
$\omega_{r}=\hbar k^{2}/2m_{a}$ is recoil frequency of an atom associated with the change in energy due to 
absorption 
or emission of a single photon. 
Last term in equation~(\ref{Ha}) describes coupling energy of field and condensate 
with coupling strength $\xi_{sm}=\frac{\omega_{c}}{L}\sqrt{\hbar/m_{sm}4\omega_{r}}$, where, 
$m_{sm}=\hslash\omega_{c}^{2}/(L^{2}NU^2_{0}\omega_{r})$ is the effective mass of side mode which shows that 
side mode formally 
behave like a mirror whose motion is driven by interacting radiation pressure.

\section{Langevin Equations}\label{sec:Langevin}

The Hamiltonian $\hat{H}_{T}$ accounts for the effects of dissipation in the intra-cavity field, damping of moving end mirror and depletion of BEC in the system via standard quantum noise operators ~\cite{Noise1991}. The total Hamiltonian $H$ leads to coupled quantum Langevin equations for position and momentum of moving end mirror and BEC, viz.,
\begin{eqnarray}
\frac{d\hat{c}}{dt}&=&\dot{\hat c}=(i\tilde{\Delta}+i\xi
\hat{q}-i\xi_{sm} \hat Q-\kappa)\hat{c}+\eta+\sqrt{2\kappa a_{in}},\label{2a}\\
\frac{d\hat{p}}{dt}&=&\dot{\hat p}=-\omega_{m}\hat{q}+\xi\hat{c}^{\dag}\hat{c}
-\gamma'_{m}\hat{p}+\hat{f}_{B},\label{2b}  \\
\frac{d\hat{q}}{dt}&=&\dot{\hat q}=\omega_{m}\hat{p},\label{2c}  \\
\frac{d\hat{P}}{dt}&=&\dot{\hat P}=-4\omega_{r}\hat{Q}-\xi_{sm}\hat{c}^{\dag}\hat{c}
-\gamma_{sm}\hat{P}+\hat{f}_{1m},\label{2d} \\
\frac{d\hat{Q}}{dt}&=&\dot{\hat Q}=4\omega_{r}\hat{P}-\gamma_{sm}\hat{Q}+\hat{f}_{2m}.\label{2e}
\end{eqnarray}
In above equations $\tilde{\Delta}=\Delta _{c}-NU_{0}/2$, whereas $\hat{a}_{\mathrm{in}}$ is Markovian input noise of the 
cavity field. The term $\gamma'_{m}$ gives mechanical energy decay rate of the moving end mirror and 
$\hat{f}_{B}$ is Brownian noise operator~\cite{Pater06, Vitali2001}. The term $\gamma_{\rm sm}$ represents damping of BEC due to 
harmonic trapping potential which effects momentum side modes while, $\hat{f}_{1m}$ and $\hat{f}_{2m}$ are the associated 
noise operators assumed to be Markovian.


To consider the classical dynamics, 
we consider positions and momenta as classical variables. In the steady state, we use adiabatic approximation
in which we set time derivative of the field as zero in equation(\ref{2a}). The steady state field defined as 
\begin{equation}
 \alpha=\frac{\eta}{\kappa +i(\Delta_{a}-\xi q +\xi_{sm} Q)}.
\end{equation}
Here, $\hat c\rightarrow \alpha$ is the classical limit.
We use this relation in equations~(\ref{2b},\ref{2d}) and put these equations in the second derivative of 
equations~(\ref{2c},\ref{2e}).
As a result the coupled equations of motion for the moving end-mirror and BEC reveal their coupled nonlinear 
dynamics, that is,
\begin{eqnarray}
\frac{d^{2}q}{dt^{2}}&=&-\omega_{m}^{2}q
 +\frac{\omega_{m}\xi\eta^{2}}{\kappa^{2}
+(\tilde{\triangle}+\xi q-\xi_{sm} Q)^{2}}\label{3},\\
\frac{d^{2}Q}{dt^{2}}&=&-4\omega_{r}^{2}Q
-\frac{4\omega_{r}\xi_{sm}\eta^{2}}{\kappa^{2}
+(\tilde{\triangle}+\xi q-\xi_{sm}Q)^{2}}\label{3a}.
\end{eqnarray}
Now we describe Hamiltonian as $H=K+V$ where, $K$ is kinetic energy of the system and the potential energy 
$V$ is obtained from equation (\ref{3}). We further assume that $\xi_{sm}<<\xi$ which provides 
the BEC evolution as harmonic oscillator with frequency $4\omega_{r}$, such that, 
$Q=Q_{0}\cos(4\omega_{r}t)$, where, $Q_{0}$ is the maximum displacement of BEC from mean position. 
We introduce some dimensionless parameters defined as, $\gamma=\omega_{m}/\omega_{r}$, 
$\beta=\eta^{2}/\kappa^{2}$, 
$\mu=\Delta/\kappa$, $\mu_{1}=\xi/\kappa$, $\lambda=\frac{\xi_{sm}}{\xi}Q_{0}$ where, 
$Q_{0}=Q_{1}/\lambda_{p}$ and using dimensionless time,$\tau=\omega_{r}t$. 
By using above parameters and integrating equation (\ref{3}) we get,
\begin{equation}\label{4}
V=\frac{d q}{d\tau}=\frac{1}{2}\gamma^{2}q^{2}
-\int\frac{\gamma\xi\beta/\omega_{r}}{1+[\mu+\mu_{1}\{q-\lambda\cos(4\tau)\}]^{2}}dq
\end{equation}
Now, we can write Hamiltonian $H$,
\begin{equation}\label{5}
H=-\frac{\partial^{2}}{2\partial q^{2}}+\frac{1}{2}\gamma^{2}q^{2}
-\int\frac{\gamma\xi\beta/\omega_{r}}{1+[\mu+\mu_{1}\{q-\lambda\cos(4\tau)\}]^{2}}dq
\end{equation}
By using transformation $x=Q-\lambda\cos(4\tau)$, we find an
effective Hamiltonian for the mirror, that is,
\begin{equation}\label{6}
H_{eff}=\frac{1}{2}\tilde{p}^{2}+\frac{1}{2}x^{2}+x\lambda_{eff}\cos(4\tau)
-\gamma_{m}\beta \arctan(\mu-\mu_{1}x),
\end{equation}
where, $\gamma_{m}=\frac{4\xi}{\gamma\omega_{m}}$ and $\lambda_{eff}=(1+\frac{32}{\gamma^{2}})\lambda$ 
is the effective modulation.

In our later work we consider the power of external field $P=0.0164$ mW with frequency 
$\omega_{p}=3.8\times2\pi\times10^{14}$ Hz and wave length $\lambda_{p}=780$ nm. 
Coupling of external field and intra-cavity field is $\eta=18.4\times2\pi$ MHz and frequency of intra-cavity field is considered $\omega_{c}=15.3\times2\pi\times10^{14}$ Hz 
which produces recoil of $\omega_{r}=3.8\times2\pi$ kHz in atoms placed in cavity of length $l=1.25\times10^{-4}$ m and having decay rate $\kappa=1.3\times2\pi$ MHz. 
Moving end mirror of cavity should be perfect reflector that oscillates with a frequency 
$\omega_{m}=\omega_r/k^{\hspace{-2.1mm}-}=15.2\times2\pi$ kHz. 
The  mirror-field and condensate-field coupling strengths are, respectively, $\xi=15.07$ MHz and $\xi_{sm}=14.39$ kHz. 
Detuning of the system is taken as $\Delta=\Delta_{c}+ \frac{U_{0}N}{2}=0.52\times2\pi$ MHz, where vacuum Rabi frequency 
of the system is $U_{0}=3.1\times2\pi$ MHz and number of ultra cold atoms placed in the BEC-optomechanical system are 
$N=2.8\times10^{4}$~\cite{Esslinger,Carmon05,Carmon07,AIP,Ferdinand}. These experimental parameters are used to obtain numerical 
results presented in all the figures.
\section {Dynamical localization of Mirror}\label{sec:NumDL}
In order to discuss classical dynamics, we solve Hamilton's equations
and in quantum domain, we solve time dependent Schr\"odinger equation. The classical dynamics of mirror is studied by Poincar\'e surfaces of section for different modulation amplitudes.
\begin{figure}[htp]
\includegraphics[width=8cm]{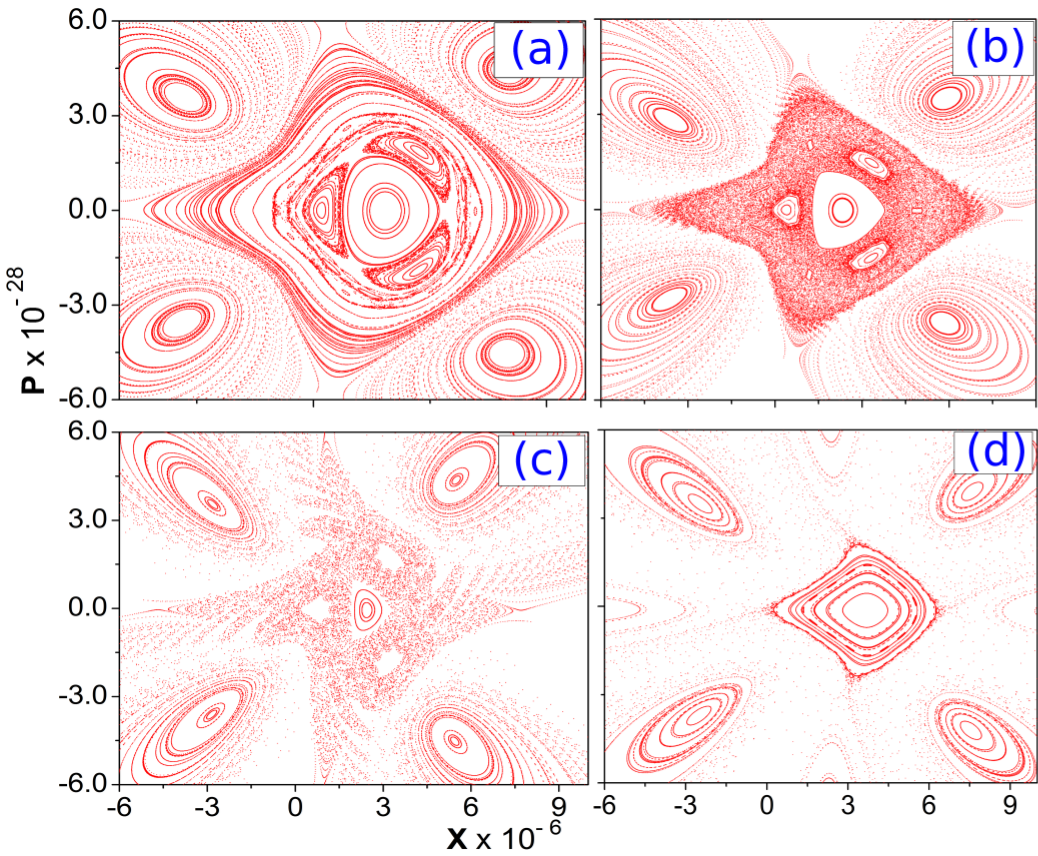}
\caption{Phase Space (Poincar\'e surface of section) plots of moving end mirror for different effective modulation 
amplitudes $\lambda_{eff}=2.6\times10^{-6}m,5.2\times10^{-6}m,1.05\times10^{-5}m,3.15\times10^{-5}m$ in figure (a), (b), (c) and (d),
respectively. Where, $\mu=-0.4$, $\mu_{1}=2$, $\beta=1.8$ and $\gamma_{m}=0.6034$.}
\label{fig:PhaseSpace}
\end{figure}
Fig-\ref{fig:PhaseSpace} shows Poincar\'e surfaces of section for $500$ initial position of the mirror.
For sufficiently large modulation classical phase space shows mixed behavior i.e. stable islands are emerged in chaotic regions.
We note that chaotic regions increase with the modulation amplitude while resonant regions shrink. As dynamical localization emerges 
from quantum evolution of wave packet in classically chaotic regime, we study the behavior of momentum and position dispersion of 
movable mirror for modulation amplitude showing mixed phase space behavior.
The quantum dynamics of the mirror is explored by evolving a Gaussian wave packet, $\Psi(x)$, 
defined at $\tau=0$,
\begin{equation}
\Psi(x)=\frac{1}{\sqrt{\sqrt{2\pi}\Delta x}}\exp(-\frac{(x-x_{0})^{2}}{2\Delta x^{2}})
\times\exp(-i\frac{p_{0}x}{k^{\hspace{-2.1mm}-}}),
\label{eq:Wavepacket}
\end{equation}
where, initial dispersion in position and momentum is defined in such a way that it satisfy the minimum uncertainty condition.
For all numerical calculations, we use $\Delta x=1$, $\Delta p=0.5$ and scaled Planck's constant is $k^{\hspace{-2.1mm}-}=1$. 
Whenever classical parameters are evaluated for comparison, we use a Gaussian ensemble with the same initial dispersion in position 
and momentum and initially placed with same mean position and momentum as in the case of quantum wave packet.
\begin{figure}[htp]
\includegraphics[height=5cm, width=7cm]{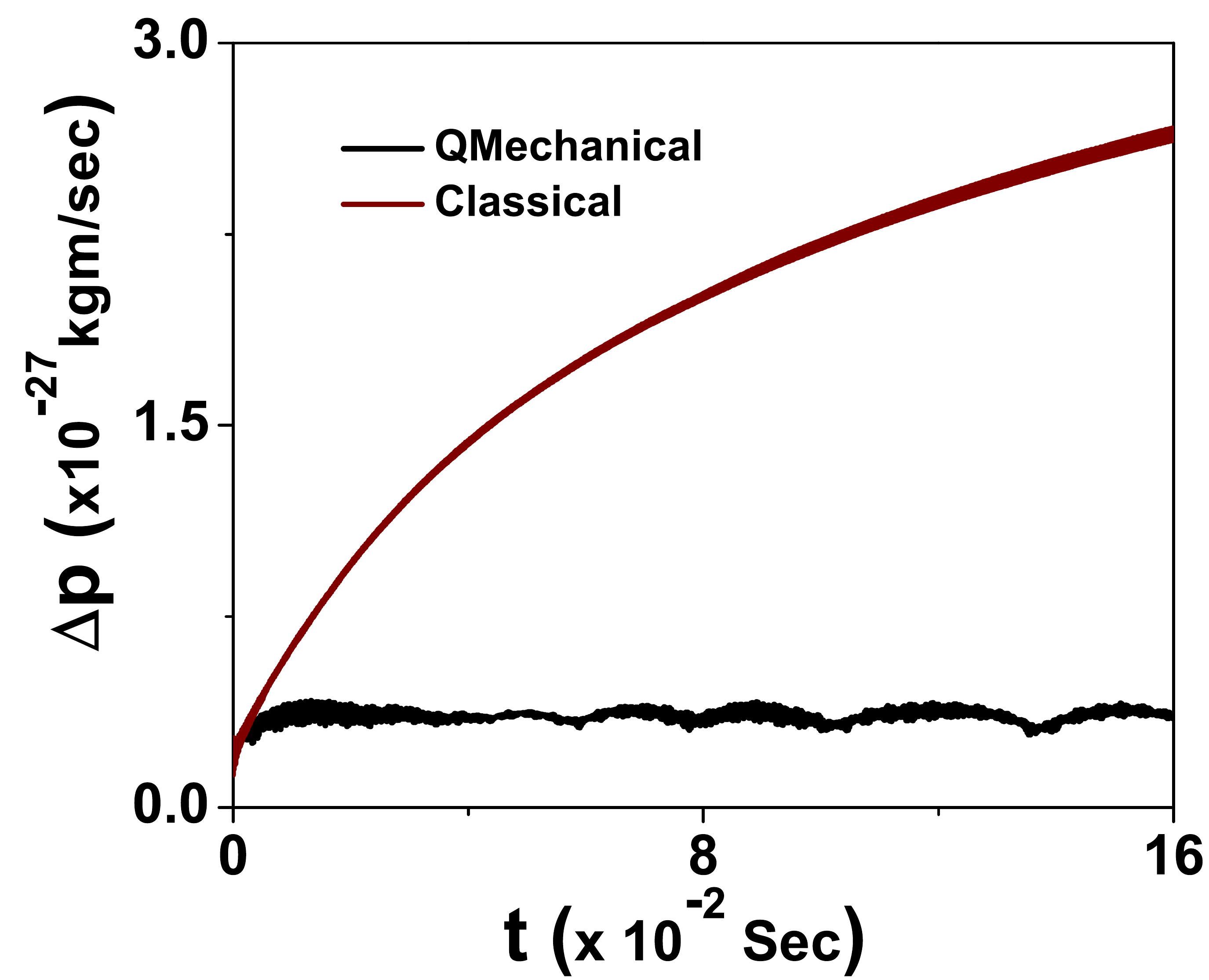}
\includegraphics[height=5cm, width=7cm]{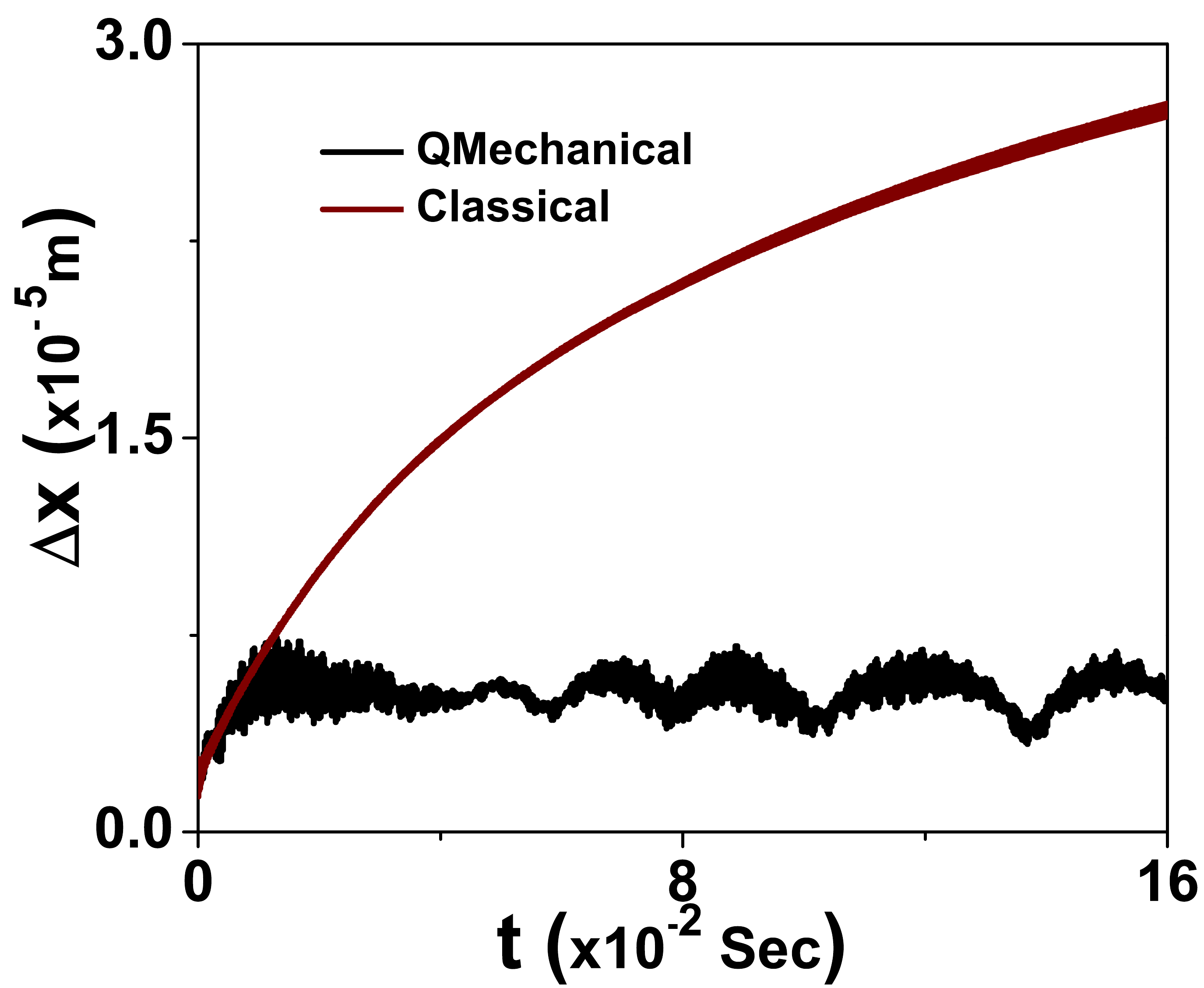}
\caption{Classical and quantum dispersion in position, $\Delta x$ (upper figure) and momentum $\Delta p$, (lower figure) 
of movable vs time at $\lambda_{eff}=1.05\times10^{-5}m$. The remaining parameters are the same as in 
Fig-\ref{fig:PhaseSpace}.}
\label{DispersionVsTime}
\end{figure}
Fig~\ref{DispersionVsTime} shows classical and quantum dispersion in position and momentum space as a function of time. 
The dispersion in momentum  space follow classical diffusion for quantum break time $t\sim 12.7 ~msec$ and after this 
time $t>12.7 ~msec$, the quantum mechanical dispersion of mirror in momentum space becomes saturated and 
oscillate around a mean value. While classical dispersion increases continuously with time, following
$ t^\alpha$ law, where $\alpha=0.6$ showing anomalous diffusion in classical domain.
Similarly, quantum dispersion in position space initially diffuses for quantum break time $t\sim 23.7~ msec$ and later saturates and show small oscillations around mean value whereas, classical dispersion in position space follows a continues diffusion even after the quantum break time. The suppression in quantum dispersion in position and momentum space is a signature of dynamical localization of mirror in position and momentum space \cite{SchliechPRL1997}.   
\begin{figure}[tbp]
\includegraphics[height=5cm, width=7cm]{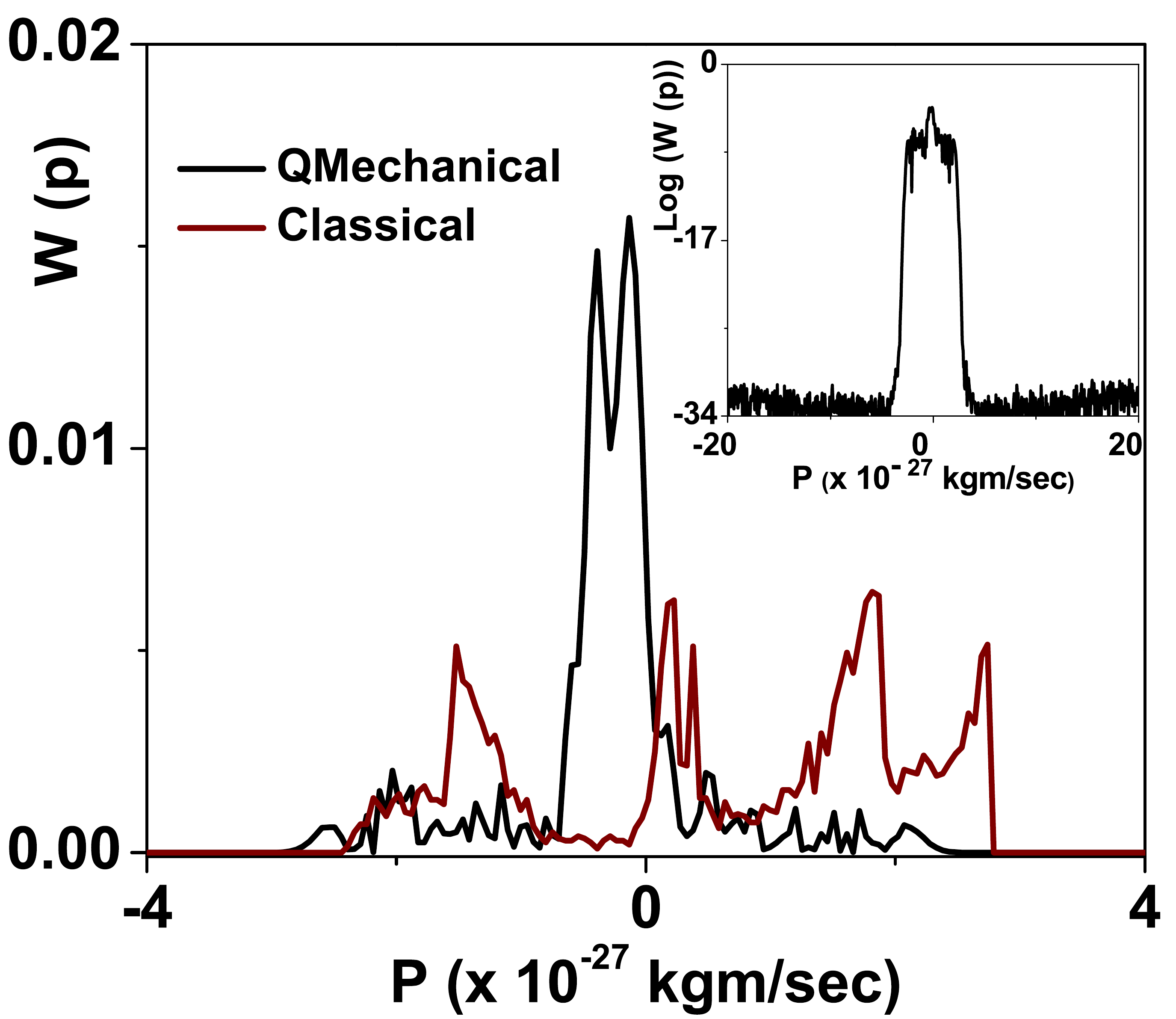}\\
\includegraphics[height=5cm, width=7cm]{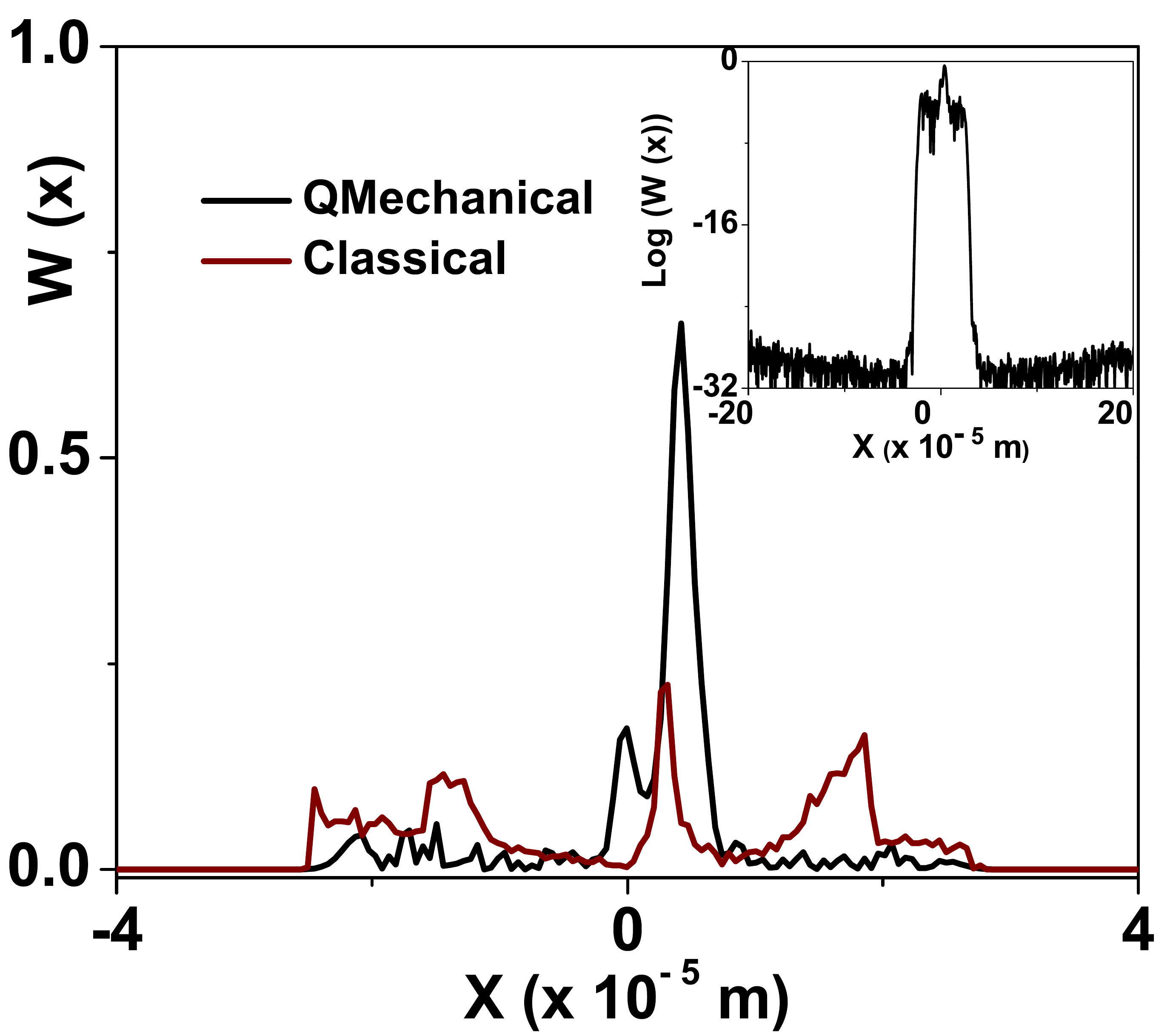}
\caption{Classical and quantum mechanical momentum distribution at time $t=131.5~msec$ in momentum space $W(p)$ (upper figure) 
and position space $W(x)$ (lower figure). The insets show exponential localization behaviour in distribution on logarithm scale. 
The remaining parameters are the same as in Fig-\ref{fig:PhaseSpace}.}
\label{fig:ProbDistribution}
\end{figure}
We study position and momentum distribution of the wave packet for $t=131.5~msec$ which is a time at which dispersion in position and momentum is dynamically localized. Fig-\ref{fig:ProbDistribution} shows classical and 
quantum mechanical time averaged probability distribution of mirror both in position, $W(x)=|\Psi(x)|^{2}$ and momentum space $W(p)=|\Psi(p)|^{2}$.
The inset show the quantum results in log scale. These results show that classical probability distribution in position and momentum space is broader 
than the quantum mechanical distribution. Classical probability distribution in position and momentum space is almost equally distributed over the entire space. While, quantum mechanical probability distribution both in position and momentum space is maximum near $x=1$. The peaks appearing in quantum mechanically position and momentum distribution  are due to stable islands in the chaotic sea as shown in Fig-\ref{fig:PhaseSpace}. The quantum mechanical distribution of wave packet in position and momentum space at large time $t=131.5~msec$ is confirmation of dynamical localization of the mirror.
The spatiotemporal behavior of mirror for $\lambda_{eff}=4$ is shown in Fig-\ref{fig:SpatioTemp} 
for the same parameters as 
in Fig-\ref{DispersionVsTime}.
The time evolution of probability distribution of quantum wave packet shows localization effects both in position 
and momentum beyond quantum 
break time. The quantum probability distribution in position and momentum space remains
localized with small fluctuations in position and momentum
space while, in classical domain distributions show ever spreading behavior in position and momentum space.
\begin{figure}[htp]
\includegraphics[width=9cm]{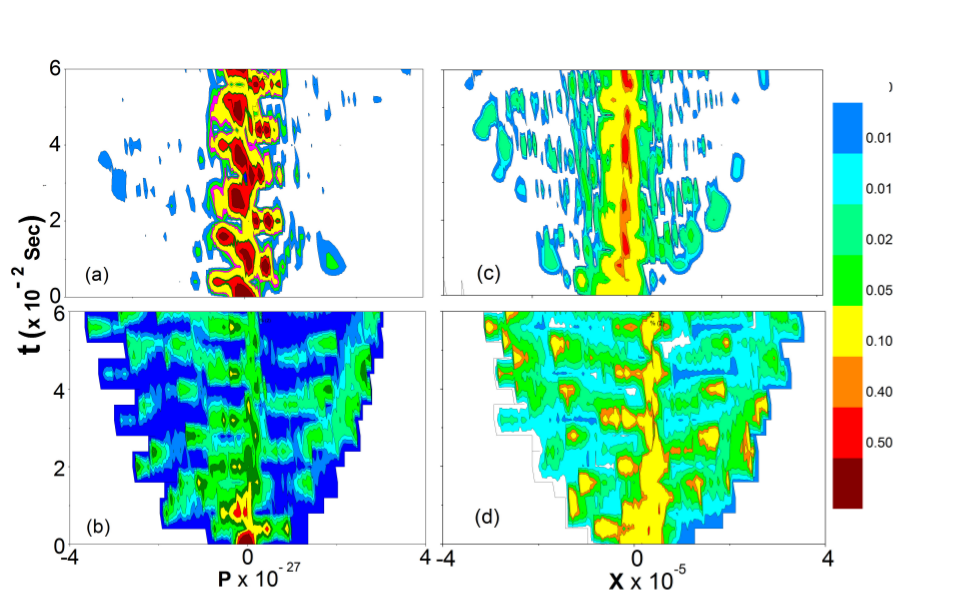}
\caption{Spatio-temporal dynamics of mirror in momentum and position space vs time shows:
(a) quantum mechanical probability distribution in momentum space, (b) classical distribution in momentum space,
(c) quantum mechanical probability distribution in position space, and (d) classical distribution in position space.
The remaining parameters are the same as in 
Fig-\ref{fig:PhaseSpace}.}
\label{fig:SpatioTemp}
\end{figure}
\section{Effects of modulation}\label{sec:ModEffect}

\begin{figure}[tbp]
\includegraphics[height=5cm, width=7cm]{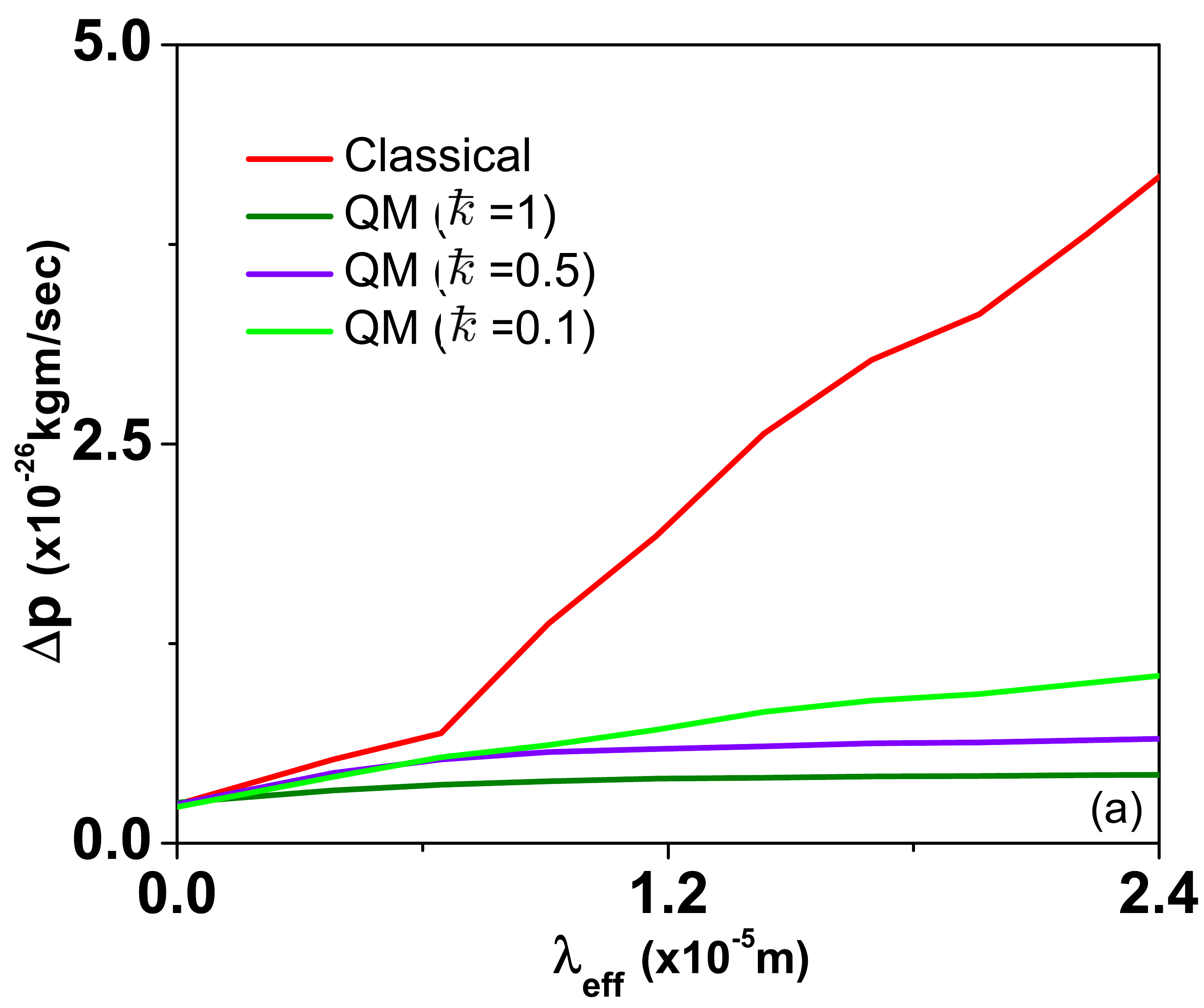}
\includegraphics[height=5cm, width=7cm]{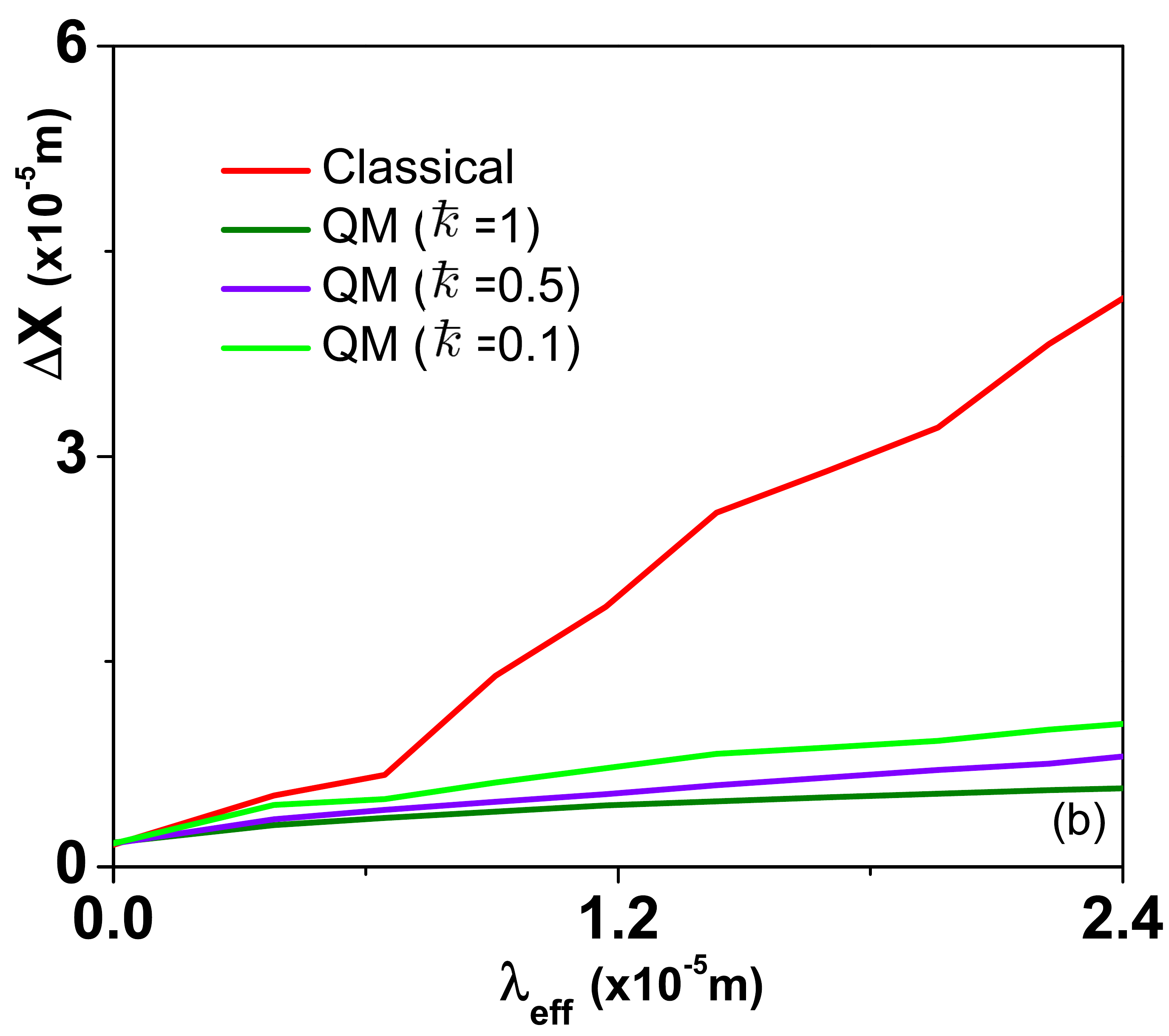}
\caption{Dispersion in momentum and position space vs modulation at $t=131.5 ~msec$.
Upper figure shows the behavior of momentum dispersion vs effective modulation amplitude $\lambda_{eff}$ while 
lower figure 
shows the behavior of position dispersion vs  $\lambda_{eff}$. Quantum dispersion behavior is plotted for  
rescaled Planck's constant $k^{\hspace{-2.1mm}-}=0.1, 0.5, 1$.
The remaining parameters are the same as in Fig.~\ref{fig:PhaseSpace}}
\label{fig:DispersionVs}
\end{figure}

 In this section, we study the effects of modulation on the position and momentum dispersion of movable mirror.
Fig-\ref{fig:DispersionVs} shows effects of modulation on classical and quantum dispersion of the mirror in position 
and momentum space. The classical ensemble or quantum wave packet is evolved for $t=131.5~msec$ for 
fixed rescaled Planck's constant $k^{\hspace{-2.1mm}-}=0.1,~1$ and dispersion at $t=131.5 ~msec$ 
is plotted 
versus fixed effective modulation.
We observe that when there is no modulation, classical and momentum dispersions show not a significant
difference both in position and momentum space. But as effective modulation increases, 
the quantum dispersion shows dynamical localization effects while classical dispersion in position 
and momentum shows 
diffusion. 
The localization effects appear for the effective modulation where corresponding classical phase space is mixed.
For small effective modulation, corresponding phase space is near integrable and no dynamical 
localization effects appear while 
for large $\lambda_{eff}$, localization effects are seen. 
We get strong dynamical localization at that particular effective
modulations where the difference between classical dispersion and quantum mechanical dispersion is maximum 
and and vice 
versa is also true.

The Fig-\ref{fig:DispersionVs} also shows behavior of quantum dispersion in position and momentum 
space for different values 
of effective Planck's constant, $k^{\hspace{-2.1mm}-}$. As $k^{\hspace{-2.1mm}-}$ is decreased curve 
representing 
quantum dispersion approaches to the curve showing classical dispersion which is a manifestation of correspondence principle.
\section{Conclusion}
We conclude that the motion of moving end mirror weakly coupled with Bose-Einstein condensate in the optomechanical cavity (Fabry-Perot Cavity) shows dynamical localization in position and momentum space. Moreover, for increasing modulation dynamical localization in momentum space doesn't disappear as in the case of driven optical lattice \cite{Bardroff1995} where this localization effect is governed by Bessel function with modulation strength as argument of Bessel function and dynamical localization revives for the value of arguments where Bessel function is zero. In position space, localization length slowly increases with modulation as shown in the Fig-\ref{fig:DispersionVs}.
The set of parameters used here makes this phenomenon experimentally feasible.

\end{document}